\documentclass[letterpaper]{article} 
\usepackage{aaai24}  
\usepackage{times}  
\usepackage{helvet}  
\usepackage{courier}  
\usepackage[hyphens]{url}  
\usepackage{graphicx} 
\usepackage{natbib}  
\usepackage{caption} 
\usepackage{algorithm}
\usepackage{algorithmic}
\usepackage{svg}
\usepackage{amsmath}
\usepackage{multirow}
\usepackage{bm}

\usepackage{booktabs} 
\usepackage{graphics}
\usepackage{array}
\usepackage[english]{babel}
\usepackage[T1]{fontenc}
\usepackage{textcomp}
\usepackage{latexsym}
\usepackage{natbib}
\usepackage[labelformat=simple]{subcaption}
\usepackage{algorithm}
\usepackage{algorithmic}
\usepackage{makecell}

\usepackage{newfloat}
\usepackage{listings}
\DeclareCaptionStyle{ruled}{labelfont=normalfont,labelsep=colon,strut=off} 
\floatstyle{ruled}
\newfloat{listing}{tb}{lst}{}
\floatname{listing}{Listing}
\title{UniGen: A Unified Generative Framework for Retrieval and Question\\Answering with Large Language Models}

\author{
    Xiaoxi Li\equalcontrib, 
    Yujia Zhou\equalcontrib, 
    Zhicheng Dou
}
\affiliations{
    Gaoling School of Artificial Intelligence, Renmin University of China, Beijing, China\\
    Engineering Research Center of Next-Generation Intelligent Search and Recommendation, Ministry of Education\\
    \{xiaoxi\_li, zhouyujia, dou\}@ruc.edu.cn
}

\usepackage{bibentry}

\begin{document}

\maketitle

\begin{abstract}

Generative information retrieval, encompassing two major tasks of Generative Document Retrieval (GDR) and Grounded Answer Generation (GAR), has gained significant attention in the area of information retrieval and natural language processing. Existing methods for GDR and GAR rely on separate retrieval and reader modules, which hinder simultaneous optimization. To overcome this, we present \textbf{UniGen}, a \textbf{Uni}fied \textbf{Gen}erative framework for retrieval and question answering that integrates both tasks into a single generative model leveraging the capabilities of large language models. UniGen employs a shared encoder and two distinct decoders for generative retrieval and question answering. To facilitate the learning of both tasks, we introduce connectors, generated by large language models, to bridge the gaps between query inputs and generation targets, as well as between document identifiers and answers. Furthermore, we propose an iterative enhancement strategy that leverages generated answers and retrieved documents to iteratively improve both tasks. Through extensive experiments on the MS MARCO and NQ datasets, we demonstrate the effectiveness of UniGen, showcasing its superior performance in both the retrieval and the question answering tasks.

\end{abstract}

\section{Introduction}

Generative information retrieval has been a focal point of research in recent years, concerning the generation of relevant information from a vast corpus, such as Wikipedia, in response to a specific query. This field primarily encompasses two tasks: Generative Document Retrieval (GDR)~\cite{Metzler2021forum,Tay2022DSI, zhuang2022dsiqg, Wang2022nci} and Grounded Answer Generation (GAR)~\cite{guu2020retrieval,lewis2020retrieval,izacard2020leveraging}. GDR retrieves a ranked list of documents in response to a query through an encoder-decoder architecture that directly generates document identifiers (docids). Concurrently, GAR generates an answer that matches a specific segment of grounding information, in response to the user's query.

The generative information retrieval landscape has been dramatically reshaped by recent advances in GDR and GAR. For the GDR task, the seminal work of~\cite{Metzler2021forum} has been instrumental, where document retrieval is accomplished by directly generating document identifiers via end-to-end training generation models. Subsequent research has built upon this work, notably enhancing the indexing strategy~\cite{Tay2022DSI, zhuang2022dsiqg, Wang2022nci}, identifier design~\cite{Tay2022DSI, Bevilacqua2022SEAL, zhou2022ultron, sun2023learning, zhang2023term, tang2023semantic}, and dynamic corpora~\cite{mehta2022dsi++}. For the GAR task, prevalent models such as REALM~\cite{guu2020retrieval}, RAG~\cite{lewis2020retrieval}, FID~\cite{izacard2020leveraging}, $\text{EMDR}^2$~\cite{singh2021end} and Atlas~\cite{izacard2022few} have employed dense retrieval models to retrieve relevant documents, which are then synthesized by generative models to yield the final answer.

\begin{figure}[t]
	\centering
	\setlength{\abovecaptionskip}{0.3cm}
	\includegraphics[width=1\linewidth]{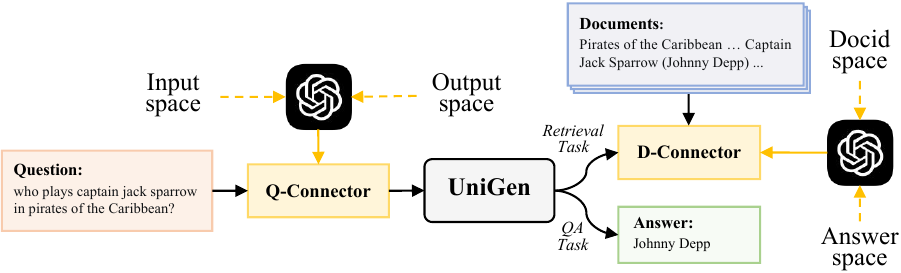}
	\caption{Illustration of the unified generative framework, which combines retrieval and question answering tasks through LLM-generated connectors.
    }
	\label{fig:first}
\end{figure}

Despite the advancements, optimizing generative retrieval and question answering (QA) tasks individually requires separate training techniques, distinct training data, and additional time costs. To address these challenges, we propose to utilize a single model to optimize both tasks simultaneously. Noting that both tasks could employ an encoder-decoder structure and possess two essential characteristics: (1) the need for a profound comprehension of the semantic significance behind the query input, and (2) the necessity to comprehend and memorize knowledge in the corpus. Drawing inspiration from these shared characteristics, we propose using a unified framework that is capable of jointly generating docids and answers, facilitating knowledge sharing, and ultimately reinforcing performance on downstream tasks.

More specifically, we propose \textbf{UniGen}, a \textbf{Uni}fied \textbf{Gen}erative framework that enhances retrieval and QA tasks concurrently. UniGen employs a \textbf{shared encoder} and two \textbf{distinct decoders}: the retrieval decoder and the QA decoder. By leveraging a shared encoder, we can improve the model's comprehension of the input through shared knowledge from both tasks, resulting in enhanced overall performance. As shown in Figure~\ref{fig:first}, the retrieval decoder generates docids for the retrieval task, while the QA decoder generates answers for the QA task. By utilizing such a shared encoder and separate decoders, our proposed UniGen framework enhances the comprehension of model input resulting in improved performance of both tasks.

Nevertheless, there are two notable gaps in such a unified generative IR framework that hinder the training process, which include: (1) the input-output gap: The input queries are often brief and lack contextual semantics, leading to a disparity between the query inputs and the generation targets. (2) The docid-answer gap: Conventional docids are typically unreadable sequences, posing a challenge to learn jointly with answer generation and thus creating a gap between docids and answers. To address these gaps, we introduce the concept of \textbf{Connectors} to serve as bridges. Specifically, we introduce the Q-Connector and the D-Connector, which enrich the query's context and refine the document's content, thereby bridging the input-output gap and the docid-answer gap, respectively. Considering generating these connectors is a highly knowledge-intensive task, we propose leveraging large language models (LLMs), which have recently gained significant attention \cite{touvron2023llama, chiang2023vicuna, chowdhery2022palm}, to effectively accomplish this task. Figure~\ref{fig:first} illustrates this approach.

Furthermore, previous works \cite{lewis2020retrieval, mao2020generation} have shown that the integration of the retrieval and QA tasks allows for a mutually beneficial relationship. Specifically, the documents acquired through the retrieval decoder serve as supplementary knowledge to enhance answer generation. Simultaneously, the answers generated by the QA decoder can contribute to more effective document retrieval. Building upon this insight, we further propose an \textbf{Iterative Enhancement Strategy} to optimize the performance of both retrieval and QA tasks at the data level. This strategy entails utilizing the retrieved documents and generated answers from the previous iteration as inputs for subsequent model iterations. We continuously refine the model input by employing this iterative process, resulting in superior performance in both tasks.

A series of experiments conducted on the public datasets {MS MARCO Question Answering}~\cite{nguyen2016ms} and {Natural Questions (NQ)}~\cite{kwiatkowski2019natural} validate the effectiveness of our proposed methods. The results demonstrate significant improvements in both retrieval and QA performance compared to baseline models. 

The paper makes the following key contributions:
\begin{itemize}
\item \textbf{Unified Generative Framework:} We develop a generative framework that incorporates a multi-decoder structure to simultaneously learn retrieval and QA tasks.
\item \textbf{LLM-enhanced Connectors:} We introduce Q-connector and D-connector generated by LLMs, which establish semantic connections in the input-output and docid-answer spaces, enhancing query semantics and refining document content, respectively.
\item \textbf{Iterative Enhancement Strategy:} We propose an iterative approach to improve both generative retrieval and QA tasks by leveraging the generated answers and the retrieved documents.
\end{itemize}

\begin{figure*}[!t]
	\centering
	\vspace{-0.1cm}
	\setlength{\abovecaptionskip}{0.25cm}
	\includegraphics[width=0.8\linewidth]{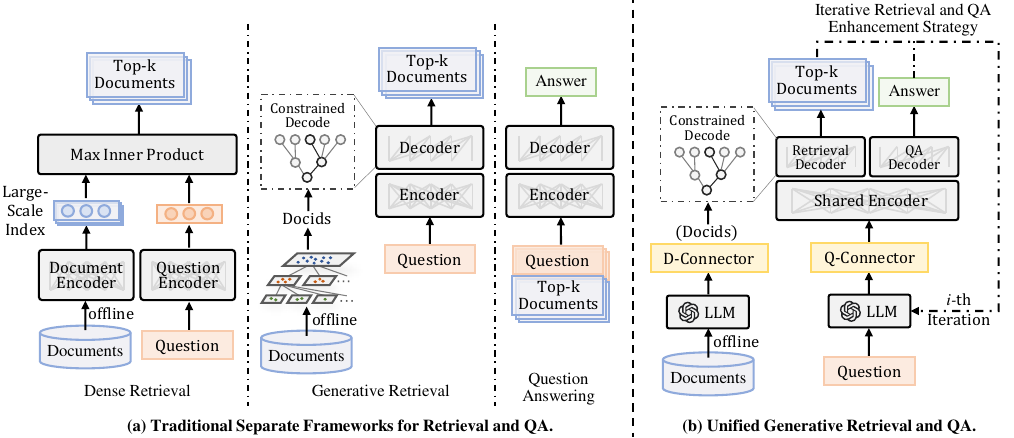}
	\caption{The comparison between traditional separate frameworks and our unified framework for retrieval and QA. (a) Traditional approaches typically employ separate and independently designed structures for retrieval and QA tasks. (b) Our proposed framework incorporates a multi-decoder structure to simultaneously achieve retrieval and QA tasks in a generative manner. To effectively enhance the performance of both tasks, we introduce LLM-generated Q-connector and D-connector, along with an iterative enhancement strategy.}
	\label{fig:model}
\end{figure*}

\section{Related Work}

\subsubsection{Generative Retrieval.} 
Generative retrieval is an innovative approach to information retrieval that leverages the parameters of pre-trained language models as differentiable indices~\cite{Tay2022DSI}, enabling the direct generation of relevant document identifiers. Recent research in this field primarily focuses on document representation and model training.
For document representation, existing studies draw inspiration from DSI~\cite{Tay2022DSI} and explore various approaches such as atomic identifiers, text fragments, and semantic clusters. Among these, text fragments stand out due to their ease of use and interpretability. For instance, Ultron~\cite{zhou2022ultron} utilizes the document URL and title as representations, while SEAL~\cite{Bevilacqua2022SEAL} considers all n-grams within a file as potential identifiers. MINDER~\cite{li2023multiview} takes a multi-view approach, incorporating synthetic identifiers, titles, and substrings.
For model training, a simple yet effective method involves using generated pseudo-query data to train the model to learn the mapping between pseudo-queries and their corresponding docids~\cite{zhuang2022dsiqg, Wang2022nci, zhou2023genrrl, wang2023novo, Zhou2022DynamicRetriever}. Subsequently, labeled query-docid data is employed to further refine the model. Another notable contribution is TOME~\cite{ren2023tome}, which proposes a two-stage model structure that first generates a paragraph relevant to the query and then generates the URL associated with the paragraph.

\subsubsection{Open-Domain Question Answering.} 
Open-domain question answering refers to providing solutions to queries without depending on contextual information. It involves two primary forms: closed-book and open-book.
In closed-book QA, models cannot access external knowledge banks and must internalize all necessary information within their parameters. Earlier works such as T5~\cite{raffel2020exploring}, BART~\cite{lewis2019bart}, and GPT~\cite{brown2020language} attempt closed-book QA by pre-training on massive text corpora, but still struggle with knowledge-intensive questions. 
In open-book QA, models can utilize knowledge bases like Wikipedia during answer generation. The typical process involves two main components: a retrieval module that searches knowledge bases for relevant contexts, and a reading module that analyzes the retrieved information to formulate a solution. For example, popular models like DPR~\cite{karpukhin2020dense}, RAG~\cite{lewis2020retrieval}, and EMDR$^2$~\cite{singh2021end} employ a dual-encoder dense retriever built upon BERT~\cite{devlin2018bert}, along with another BERT-based model for answer extraction or a T5/BART-based model for answer generation. Large language models (LLMs) have recently shown promising results in open-domain QA~\cite{yu2022generate, sun2022recitation, ram2023context, shi2023replug, borgeaud2022improving, liu2023reta}. For instance, GenRead~\cite{yu2022generate} prompts an LLM to generate context documents instead of using a retriever. Combining generation and retrieval techniques can further improve performance. RECITE~\cite{sun2022recitation} suggests asking the LLM to generate support paragraphs containing the answer, which are then used as an additional prompt along with the question. 


\section{Methodology}

In this section, we present a complete overview of our proposed framework, which aims to tackle generative retrieval and QA tasks. We will begin by defining these tasks and then dive into the structure and training methodologies employed in our unified framework.

\subsection{Task Formulation}


Consider a document $d$ in a document corpus and let $d'$ denotes the pre-built docid of document $d$.
For generative retrieval task, given a query $q$, we obtain the relevance $\mathcal{R}$ between $q$ and each document $d$ by
\begin{equation}
\begin{aligned}
\label{eq:gr}
    \mathcal{R}(q,d)
    =f_{\text{retr}}(d'|q;\theta,\phi)
    =\prod \limits_{i=1}^T f_{\text{retr}}(d'_{i}|d'_{<i},q;\theta,\phi),
\end{aligned}
\end{equation}
where $T$ is the length of the target document identifier $d'$, $d'_{i}$ is the $i_{th}$ token of $d'$, $f_{\text{retr}}$ is the generative retrieval model comprising an encoder with parameters $\theta$ and a retrieval decoder with parameters $\phi$. The model is trained to maximize the likelihood of generating the target document identifier in Eq.~(\ref{eq:gr}). Teacher forcing is used during training to optimize the following cross-entropy loss:
\begin{equation}
\label{eq:lgr}
    \mathcal{L}_{\text{retr}} = - \sum_{i=1}^T \text{log} f_{\text{retr}}(d_{i}|d'_{<i},q;\theta,\phi).
\end{equation}

Similarly, for QA task, given a query $q$, the probability $\mathcal{A}$ of generating answer $a$ is obtained by
\begin{equation}
\label{eq:qa}
    \mathcal{A}(a|q)
    =f_{\text{qa}}(a|q;\theta,\mu)
    =\prod \limits_{i=1}^{T'} f_{\text{qa}}(a_{i}|a_{<i},q;\theta,\mu),
\end{equation}
where $T'$ is the length of answer $a$, $a_{i}$ is the $i_{th}$ token of answer $a$, $f_{\text{qa}}$ is the generative QA model with a shared encoder with parameters $\theta$ and a distinct QA decoder with parameters $\mu$. Similarly, the optimization of the parameters $\theta$ and $\mu$ is achieved through the standard seq-to-seq objective, which consists of maximizing the likelihood of the target sequence in Eq.~(\ref{eq:qa}) by employing teacher forcing. The QA loss function can be represented by
\begin{equation}
\label{eq:lqa}
    \mathcal{L}_{\text{qa}} = - \sum_{i=1}^{T'} \text{log} f_{\text{qa}}(a_{i}|a_{<i},q;\theta,\mu).
\end{equation}

\subsection{UniGen: Unified Generative Retrieval and QA}

This section discusses the details of the UniGen framework proposed in the paper, including the overall model structure, LLM-based connectors generation, joint learning method, and iterative enhancement strategy. 

\subsubsection{Model Architecture}

Our proposed UniGen framework introduces a multi-decoder structure to simultaneously tackle the retrieval and the QA tasks. This is different from conventional methods that depend on separate and independently designed architectures for each task. Figure~\ref{fig:model} demonstrates the contrast between our UniGen framework and the traditional methods, where dense retrieval relies on large-scale document indices, and generative retrieval and QA methods are usually distinct modules.

The architecture of our model comprises an encoder and two separate decoder heads: a retrieval decoder and a QA decoder. The encoder takes the enhanced query generated by the LLM as input, denoted by Q-Connector. The retrieval decoder employs constrained beam search within a prefix tree to generate a ranked list of document identifiers. These identifiers are represented by the connector generated by the LLM from the document side, represented by the D-Connector. At the same time, the QA decoder generates the answer text. By using a joint architecture for retrieval and QA, our model optimizes both tasks simultaneously, resulting in enhanced overall system performance.

\subsubsection{LLM-based Connectors Generation}

\begin{figure}[!t]
	\centering
	\vspace{0.1cm}
	\setlength{\abovecaptionskip}{0.25cm}
	\includegraphics[width=0.98\linewidth]{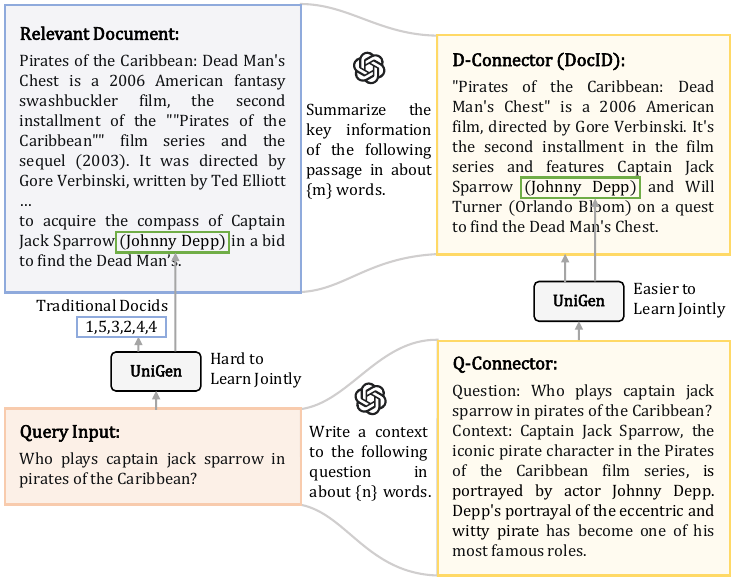}
	\caption{An example of generating LLM-based connectors from the query side and document side, with the labeled answer highlighted in the green box.}
	\label{fig:connector}
\end{figure}

Learning to generate docids and answers concurrently based on query inputs is a challenging task. Query inputs are typically short and lack context, while documents are long and contain redundant information. Directly mapping queries to documents and answers is difficult. Additionally, existing docid representations are often meaningless sequences, hindering the joint learning of generative retrieval and QA tasks. To address these issues, we propose using LLM to generate Q-Connectors and D-Connectors on the query and document sides. These connectors serve as bridges between query inputs, documents, and answer outputs. Figure~\ref{fig:connector} provides an example of the LLM-generated connectors.

Firstly, for D-Connector generation, the LLM takes the prompt of “\texttt{Summarize the key information of the following document in about \{$m$\} words.\textbackslash n Document:\{$d$\}}" along with a document $d$ as input and outputs a summary of the document called a D-Connector $d_c$.
The D-Connector serves as a docid of the document that captures its essential information, which greatly reduces the difficulty of the model's memory for long documents. Additionally, since the answer is typically a short phrase or sentence, it is easier to jointly learn with the answer generation task using the unified framework proposed in this paper.


Secondly, for Q-Connector generation, the LLM takes the prompt of “\texttt{Write a context to the following question in about \{$n$\} words.\textbackslash n Question:\{$q$\}}" and a question $q$ as input and generates a Q-Connector $q_c$.
The Q-Connector provides a contextual representation of the query, which aids in generating relevant docids and accurate answers. The Q-Connector enables the model to better understand the query and its related context, thereby enabling it to effectively map to relevant docids and provide contextual knowledge for the QA task. This approach does not rely on external corpora and can achieve impressive results for QA.

\subsubsection{Joint Learning of Retrieval and QA}

Taking the Q-Connector $q_c$ as the model input, we establish the relevance between the query $q$ and each document $d$ in the set $\mathcal D$ and the probability of generating an answer $a$ by $f_{\text{retr}}(d_c|q_c;\theta,\phi)$ and $f_{\text{qa}}(a|q_c;\theta,\mu)$, respectively. Here, $\theta$, $\phi$, and $\mu$ represent the parameters of the model encoder, retrieval decoder, and QA decoder, respectively. We can modify the retrieval and QA losses from Eq.~(\ref{eq:lgr}) and Eq.~(\ref{eq:lqa}) as follows:
\begin{align}
\label{eq:loss_retr}
    \mathcal{L}_{\text{retr}}^{'} &= - \sum_{i} \text{log} f_{\text{retr}}(d_{c_i}|d_{c_{<i}},q_c;\theta,\phi),\\
\label{eq:loss_qa}
    \mathcal{L}_{\text{qa}}^{'} &= - \sum_{i} \text{log} f_{\text{qa}}(a_{i}|a_{<i},q_c;\theta,\mu),
\end{align}
where $d_{c_i}$ and $a_{i}$ denote the $i_{th}$ token in the generation of $d_c$ and $a$, respectively.

To equip the model with initial generative capabilities for both tasks, we begin by training it on synthetic training data. Previous research has demonstrated that using synthetic data can enhance the effectiveness of generative retrieval and question answering~\cite{zhuang2022dsiqg, puri2020training}. Hence, we present a two-stage training approach, including a pre-training stage and a fine-tuning stage:

In the \textbf{pre-training stage}, for each document $d$, we employ the DocT5query~\cite{nogueira2019doc2query} model to generate $K$ pseudo queries $q_k$, where $k \in \{1,...,K\}$. Next, we feed each pseudo query $q_k$ and its corresponding document $d$ into the large language model LLaMA-13B-Chat~\cite{touvron2023llama2} to generate label answers $a_k$. To simulate the Q-Connector $q_c$ generated by the LLM, we concatenate $q_k$ and $d$ as the input of our generative model, denoting $q_k+d$. This approach allows us to generate $K$ pairs of retrieval and QA training data <$q_k+d,d_c$> and <$q_k+d,a_k$> for each document $d$. 

In the \textbf{fine-tuning stage}, we proceed by training the model based on labeled <$q_c,d_c$> and <$q_c,a$> data, where $q_c$ is generated by LLM from query $q$. 

To optimize the model for both generative retrieval and QA tasks, our UniGen framework employs both the generative retrieval loss and the QA loss, denoted by Eq.~(\ref{eq:loss_retr}) and Eq.~(\ref{eq:loss_qa}), respectively. To jointly optimize the encoder parameters $\theta$, retrieval decoder parameters $\phi$, and QA decoder parameters $\mu$, we combine these two losses into a single overall loss:
\begin{equation}
\label{eq:loss}
\mathcal{L} = \lambda \mathcal{L}_{\text{retr}}^{'} + (1-\lambda) \mathcal{L}_{\text{qa}}^{'},
\end{equation}
where $\lambda$ is the regularization weight. By following this training process and optimizing the loss function as described, the model can effectively learn both retrieval and QA tasks simultaneously. We refer to this foundational model as \textbf{UniGen-Base}.

\subsubsection{Iterative Enhancement Strategy}

To further enhance the retrieval and QA performance of the model at the data level, we propose an iterative enhancement strategy. The objective is to utilize the retrieved documents and generated answers from the previous iteration as inputs for the next round of the model, as shown in the dashed portion in Figure \ref{fig:model}(c).

In each iteration, we input the top-k documents, answer, and query from the previous round into a large language model. The aim is to generate a higher-quality Q-Connector, denoted as $q_c$, and increase the likelihood of the model producing the correct answer and retrieving more relevant documents. To accomplish this, we use the following prompt: "\texttt{Given the following potentially relevant documents and the potentially correct answer, please provide the context for the question in \{$n$\} words. \textbackslash n Document:\{$d$\} \textbackslash n Answer:\{$a$\} \textbackslash n Question:\{$q$}\}". The parameter $n$ allows us to control the length of $q_c$. 

Through this iterative approach, our goal is to continuously refine the model's performance in retrieving and answering questions, ultimately improving its overall effectiveness. To strike a balance between model performance and efficiency, we have created an enhanced version of the model called \textbf{UniGen-Iter}, which incorporates two iterations on top of the UniGen-Base.

\section{Experimental Settings}

\subsection{Datasets}
To thoroughly evaluate the retrieval and question answering performance of our proposed model, we utilize two well-known datasets: MS MARCO and Natural Questions.

\textbf{MS MARCO Question Answering}~\cite{nguyen2016ms} is designed to train and test systems that can effectively generate the most probable answer given a real-world user query. We use the QnA (v2.1) dataset and extract passages from the corpus that contain labeled data, resulting in a substantial collection of approximately 100k passages and a set of 94,871 training query-answer-relevant document triplets.

\textbf{Natural Questions (NQ)}~\cite{kwiatkowski2019natural} consists of questions sampled from the Google search engine. Following the methodologies proposed by~\cite{karpukhin2020dense}, we divide each Wikipedia article into non-overlapping chunks of 100 words. To ensure a robust evaluation, we identify passages in the corpus that contain labeled data based on the training set. This meticulous process results in a diverse collection of around 100k passages and 38,191 training query-answer-relevant document triplets.

\subsection{Baselines}
We choose several baseline models for the retrieval and QA tasks, categorized into different classes.

For the retrieval task, we select three classes of models. The first class consists of \textbf{Sparse Retrieval} models, which include BM25~\cite{robertson2009probabilistic} and DocT5Query~\cite{nogueira2019doc2query}. The second class comprises \textbf{Dense Retrieval} models, such as DPR~\cite{karpukhin2020dense} and ANCE~\cite{xiong2020approximate}. Lastly, the \textbf{Generative Retrieval} models class includes DSI~\cite{Tay2022DSI}, DSI-QG~\cite{zhuang2022dsiqg}, NCI~\cite{Wang2022nci}, and Ultron~\cite{zhou2022ultron}.

Regarding the QA task, we consider three types of baseline models. The first type is \textbf{Closed-book Generation} models, represented by T5~\cite{raffel2020exploring} and BART~\cite{lewis2019bart}. The second type is \textbf{Retrieval-augmented Generation} models, which incorporate RAG~\cite{lewis2020retrieval} and a combination model that utilizes DPR, NCI, Ultron, and Fusions-in-Decoder~\cite{izacard2020leveraging}. The last type is \textbf{LLM-based Generation} models, where we directly evaluate the QA performance of gpt-3.5-turbo-0613 and LLaMA2-13B-Chat~\cite{touvron2023llama2}.

\subsection{Evaluation Metrics}
Retrieval models are evaluated using MRR and recall, which measure the average rank of the first relevant document and the proportion of relevant documents retrieved, respectively.

For QA evaluation, we use BLEU-1 (B-1) and ROUGE-L (R-L) metrics on MS MARCO. B-1 measures uni-gram overlap, while R-L measures the longest common sub-sequence overlap. On the NQ dataset, we utilize the Exact Match (EM) and F1 score, which measure exact matches and the harmonic mean of precision and recall, respectively.

\begin{table*}[!ht]
    \centering
    \small
    \setlength{\abovecaptionskip}{0.2cm}
    \caption{Overall Retrieval Performance, where \# Param indicates the size of model parameters. The best results among all experiments are emphasized in \textbf{bold}, while the best results of baseline models are \underline{underlined}. The symbol "$\dagger$" signifies that our basic model achieved superior results among all baselines in a statistically significant manner (t-test, $p < 0.05$).}
    
    \begin{tabular}{p{0.16\linewidth}|p{0.065\linewidth}<{\centering}|p{0.055\linewidth}<{\centering}p{0.055\linewidth}<{\centering}p{0.055\linewidth}<{\centering}p{0.065\linewidth}<{\centering}|p{0.055\linewidth}<{\centering}p{0.055\linewidth}<{\centering}p{0.055\linewidth}<{\centering}p{0.065\linewidth}<{\centering}}
        
    \toprule
        \multirow{2}[2]{*}{Model} & \multirow{2}[2]{*}{\# Params} & \multicolumn{4}{c|}{MS MARCO} & \multicolumn{4}{c}{Natural Questions (NQ)} \\
        
    \cmidrule(lr){3-10}
        & & R@1 & R@5 & R@10 & MRR@10 & R@1 & R@5 & R@10 & MRR@10 \\
        
    \midrule
        \multicolumn{10}{l}{\textit{Sparse Retrieval}} \\ 
        BM25 & - & 25.70 & 53.28 & 65.85 & 37.79 & 45.36 & 72.86 & 81.72 & 57.18  \\
        DocT5Query & - & 31.14 & 60.04 & 68.29 & 42.93 & 49.43 & 76.25 & 84.10 & 60.81 \\

    \midrule
        \multicolumn{10}{l}{\textit{Dense Retrieval}} \\ 
        DPR &   220M & 36.96 & 70.92 & 80.18 & 50.69 & 60.25 & 82.60 & 86.97 & 69.90\\
        ANCE &   220M & 37.70 & \underline{72.34} & \underline{81.52} & \underline{51.70} & 61.45 & 84.25 & 88.71 & 71.30\\
        
    \midrule
        \multicolumn{10}{l}{\textit{Generative Retrieval}} \\ 
        DSI-Semantic &  250M & 28.84 & 46.22 & 52.94 & 36.60 & 46.70 & 66.34 & 70.79 & 54.73 \\
        DSI-QG & 250M  & 35.41 & 68.38 & 73.34 & 46.48 & 59.52 & 78.35 & 81.93 & 67.94  \\
        NCI & 267M & \underline{37.89} & 72.23 & 77.39 & 49.41 & 63.00 & 84.61 & \underline{88.90} & 71.77 \\
        Ultron-PQ & 257M & 37.38 & 72.07 & 78.09 & 51.47 & \underline{63.54} & \underline{85.01} & 86.34 & \underline{72.68} \\

    \midrule
        \multicolumn{10}{l}{\textit{Unified Generative Retrieval and QA (Retrieval Decode)}} \\ 
        UniGen-Base & 367M & 38.75$^\dagger$ & 72.69$^\dagger$ & 79.07 & 52.64$^\dagger$ & 63.71$^\dagger$ & 86.39$^\dagger$ & 88.74 & 72.81$^\dagger$ \\
        UniGen-Iter & 367M & \textbf{42.34}$^\dagger$ & \textbf{75.99}$^\dagger$ & \textbf{81.85}$^\dagger$ & \textbf{56.38}$^\dagger$ & \textbf{64.92}$^\dagger$ & \textbf{88.15}$^\dagger$ & \textbf{90.01}$^\dagger$ & \textbf{74.61}$^\dagger$\\

    \bottomrule
    \end{tabular}
    \label{tab:retrieval}
\end{table*}

\subsection{Implementation Details}
In our experiments, we utilize the pre-trained T5-base encoder as the shared encoder for our model. Both the retrieval decoder and QA decoder also employ the T5-base decoder with pre-trained parameters from HuggingFace Transformers \cite{wolf2019huggingface}. We incorporate the {gpt-3.5-turbo-0613} API as the LLM in our system. To generate training data, we create 10 pseudo-queries and 10 pseudo-answers for each document. During training, we set the value of $\lambda$ to 0.6.  Our training process involves a batch size of 128, a learning rate of 5e-4, and 2k learning rate warm-up steps. During inference, we employ constrained beam search for generative retrieval decoding and greedy search for QA decoding. Due to memory and time constraints, we limit the beam size to a maximum of 10. The experiments are conducted on 4 NVIDIA RTX 3090 GPUs. Further details can be found in the Appendix. The source code will be made available upon publication for additional reference.

\section{Experimental Results}

In this section, we present the results of our experiments to evaluate the performance of the proposed unified model in both retrieval and QA tasks.

\subsection{Passage Retrieval Performances}

We evaluate the retrieval performance, and the overall results are summarized in Table~\ref{tab:retrieval}.

(1) Our proposed UniGen-Base model outperforms existing baseline models in terms of most metrics. Specifically, for the MRR@10 metric, UniGen-Base outperformed the best baseline models on the MS MARCO and NQ datasets by 1.81\% and 0.83\%, respectively. This can be attributed to the joint learning strategy of retrieval and question-answering tasks, which makes the shared encoder more robust, alleviates overfitting, and improves the understanding of query inputs. 
In addition, the connectors generated by LLM on the query and document sides serve to enrich the contextual semantics of queries and refine the corpus documents, thereby facilitating the model's learning of the mapping relationship between queries and relevant docids.

\begin{figure}[!t]
\vspace{0.2cm}
    \begin{subfigure}{.495\linewidth}
    \centering
    \includegraphics[width=\linewidth]{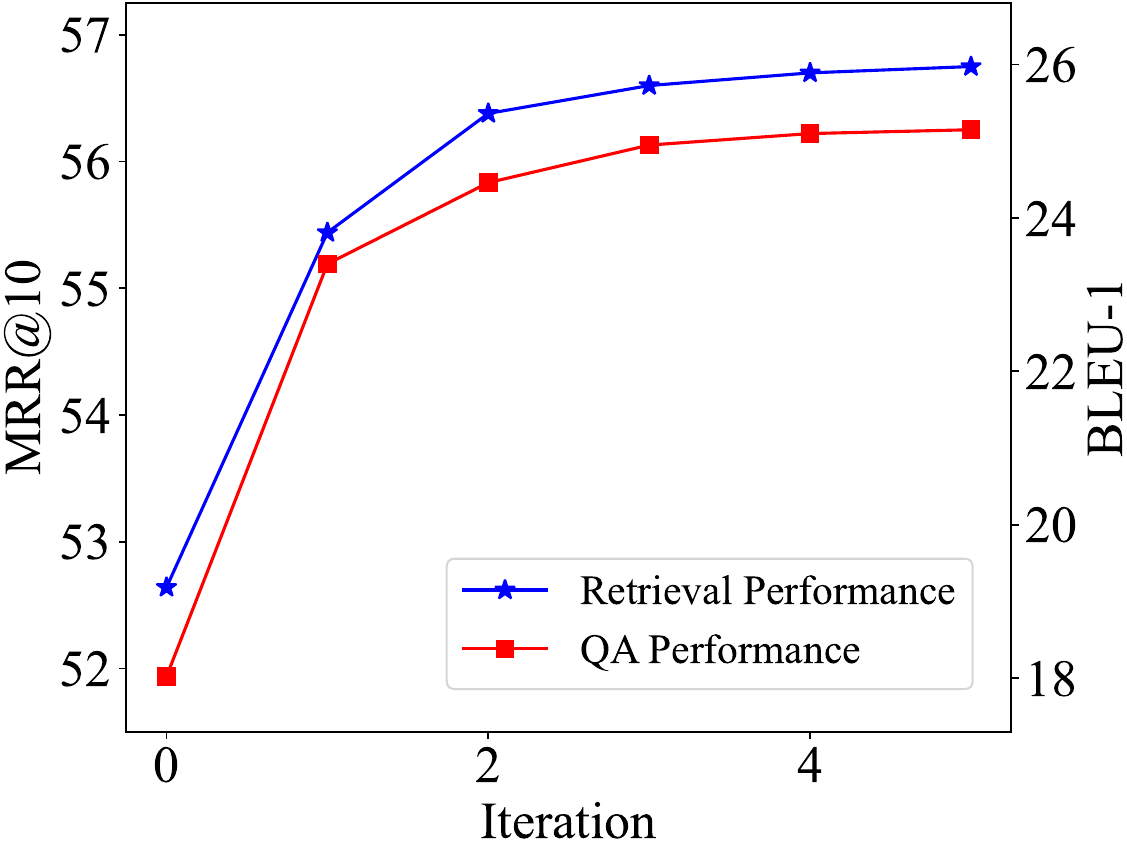}
    \caption{MS MARCO}
    \end{subfigure}
    \begin{subfigure}{.495\linewidth}
    \centering
    \includegraphics[width=\linewidth]{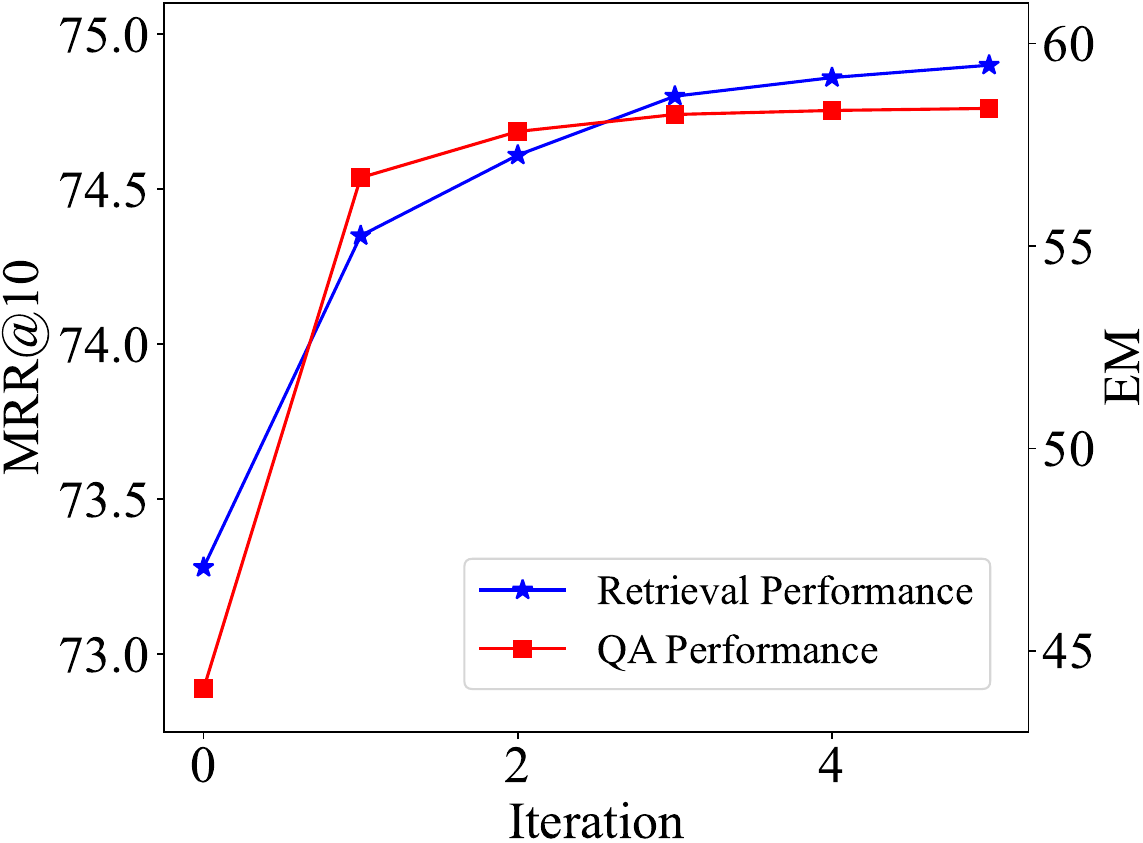}
    \caption{Natural Questions}
    \end{subfigure}
    \caption{Analysis of retrieval and QA performance with different iterations.} 
    \label{fig:iteration}
\end{figure}

(2) For the data-level unification, the proposed UniGen-Iter model achieves the best retrieval performance after two iterations, outperforming existing generative retrieval models, dense retrieval, and sparse retrieval models. Specifically, UniGen-Iter surpasses the best baseline models on the MS MARCO and NQ datasets by 11.76\% and 2.17\% in terms of R@1, respectively. Furthermore, as shown in the blue lines in Figure~\ref{fig:iteration}, a continuous improvement in retrieval performance can be observed in terms of MRR@10 on MS MARCO and NQ datasets when comparing the non-iterative approach (UniGen-Base) with the iterative methods for 1 to 5 iterations. This clearly demonstrates the effectiveness of the proposed iterative enhancement strategy in improving retrieval performance. This is because the previously retrieved documents can provide relevant external knowledge, and the generated answers can also serve as references, enabling LLM to generate more relevant Q-Connectors and continuously enhance retrieval performance over iterations.

In summary, the proposed UniGen model demonstrates superior retrieval performance compared to existing models, and the iterative enhancement strategy proves to be effective in improving retrieval performance. 

\begin{table}[!t]
    \centering
    \small
    \fontsize{8pt}{\baselineskip}\selectfont
    \caption{Overall QA Performance, where \# Param indicates the size of model parameters. The best results among all experiments are emphasized in \textbf{bold}, while the best results of baseline models are \underline{underlined}. The symbol "$\dagger$" and "$\ddagger$" signifies that our model achieved superior results in the closed-book and open-book settings, respectively.}
    
    \begin{tabular}{p{0.235\linewidth}|p{0.12\linewidth}<{\centering}|
    p{0.085\linewidth}<{\centering}p{0.085\linewidth}<{\centering}|
    p{0.085\linewidth}<{\centering}p{0.085\linewidth}<{\centering}}
        
    \toprule
        \multirow{2}[2]{*}{Model} & \multirow{2}[2]{*}{\# Params} & \multicolumn{2}{c|}{MS MARCO} & \multicolumn{2}{c}{NQ} \\
    \cmidrule(lr){3-6}
        & & B-1 & R-L & EM & F1 \\
        
    \midrule
        \multicolumn{6}{l}{\textit{Closed-book Answer Generation}} \\ 
        T5 & 220M & 14.49 & 18.74 & 30.22 & 38.40 \\
        BART & 340M & 14.83 & 19.26 & 29.76 & 37.34 \\
        
    \midrule
        \multicolumn{6}{l}{\textit{LLM-based Answer Generation (w/o finetuning)}} \\ 
        GPT-3.5 & - & 16.67 & 19.41 & 28.03 & 36.59 \\
        LLaMA2 & - & 16.08 & 18.61 & 25.27 & 33.56 \\

    \midrule
        \multicolumn{6}{l}{\textit{Retrieval-augmented Answer Generation}} \\ 
        RAG     & 540M & 17.32 & 21.83 & 53.32 & 62.86 \\
        DPR + FID & 440M & \underline{18.97} & \underline{23.15} &  54.47 & 63.53 \\
        DSI + FID & 470M & 15.64 & 19.79 & 43.58 & 52.08 \\
        Ultron + FID & 477M & 17.21 & 21.96 & \underline{55.75} & \underline{65.05} \\

    \midrule
        \multicolumn{6}{l}{\textit{Unified Generative Retrieval and QA (QA Decode)}} \\ 
        UniGen-Base  & 367M & 17.02$^\dagger$ & 21.90$^\dagger$ & 44.06$^\dagger$ & 53.39$^\dagger$  \\
        UniGen-Iter & 367M & \textbf{24.46}$^\ddagger$ & \textbf{30.31}$^\ddagger$ & \textbf{57.83}$^\ddagger$ & \textbf{67.24}$^\ddagger$ \\

    \bottomrule
    \end{tabular}
    \label{tab:qa}
\end{table}

\subsection{Question Answering Performances}

We also assess the performance of the proposed model in the QA task, and the results are shown in Table~\ref{tab:qa}. 

(1) Under the closed-book setting, where external corpora are not accessible, the model directly generates answers to input questions. Comparing the small model fine-tuned with labeled data and the large model without fine-tuning, the proposed UniGen-Base model significantly outperforms existing baseline models with statistical significance ($p < 0.05$). For the MS MARCO dataset, UniGen-Base surpasses BART by 9.10\% in terms of Bleu-1, and for the NQ dataset, it outperforms T5 by 45.80\% in terms of exact match (EM). Even without accessing external documents, UniGen-Base outperforms some retrieval-based models. This can be attributed to the Q-Connector generated by the LLM, which provides effective contextual information for query inputs. Besides, the joint learning of answer generation and D-Connector enhances the model's robustness in generating answers.
(2) Under the open-book setting, comparing with existing retrieval-augmented answer generation models, the proposed UniGen-Iter outperforms the DPR+FID model by 28.94\% in terms of Bleu-1 on the MS MARCO dataset and surpasses Ultron+FID by 3.73\% in terms of EM on the NQ dataset. 
In addition, Figure~\ref{fig:iteration} illustrates the improvement in QA performance, through the use of iterative methods compared to the non-iterative approach (UniGen-Base), as indicated by the red lines. This also highlights the effectiveness of the proposed iterative enhancement strategy in enhancing QA performance. As a result, the enhanced Q-Connectors has also contributed to a consistent enhancement in the performance of QA task throughout various iterations.

To summarize, the UniGen model outperforms existing models in QA tasks, and the iterative enhancement strategy significantly contributes to its improved performance.

\subsection{Ablation Studies}

To validate the effectiveness of our proposed unified framework for retrieval and QA, we conduct experiments where we systematically remove each module and observe the resulting performance degradation, as presented in Table~\ref{tab:ablation}.

We find that removing any of the modules, namely the shared encoder, Q-Connector, or D-Connector, leads to a noticeable decline in both retrieval and QA performance. Notably, the largest drop in performance is observed when the Q-Connector is removed. This highlights the significance of leveraging large-scale language models as external knowledge sources to provide context that is relevant to queries.
Furthermore, removing the D-Connector also has a significant impact on the final performance. This demonstrates the contribution of the D-Connector in bridging the gap between the document and the answer, surpassing the capabilities of traditional methods such as hierarchical clustering-based document identifiers.
For methods that do not utilize a shared encoder, we still observe a decrease in performance, underscoring the advantages of our unified structure. This structure enables the training of more robust encoders, resulting in improved representations of inputs and enhanced retrieval and QA performance.

\begin{table}[!t]
    \centering
    \fontsize{8.1pt}{\baselineskip}\selectfont
    \caption{Ablation study of our unified generation model on the MS MARCO dataset.}
    \begin{tabular}{p{0.34\linewidth}|p{0.095\linewidth}<{\centering}p{0.095\linewidth}<{\centering}|p{0.095\linewidth}<{\centering}p{0.095\linewidth}<{\centering}}
    \toprule
        \multirow{2}[2]{*}{Model} & \multicolumn{2}{c|}{Retrieval} & \multicolumn{2}{c}{\text{QA}} \\
    \cmidrule(lr){2-5}
        & R@1 & R@10 & B-1 & R-L \\
    \midrule
        \textbf{UniGen-Base} & \textbf{38.75} &  \textbf{79.07} & \textbf{17.02} & \textbf{21.90}  \\
        ~ w/o shared encoder & 37.89 & 78.69 & 16.49 & 21.30 \\
        ~ w/o Q-Connector & 36.14 & 77.58 & 12.32 & 15.98 \\
        ~ w/o D-Connector & 37.44 & 78.25 & 15.06 & 18.74 \\
    \midrule
        \textbf{UniGen-Iter} & \textbf{42.34} & \textbf{81.85} & \textbf{24.46} & \textbf{30.31} \\
        ~ w/o shared encoder & 41.76 & 81.29 & 23.72 & 29.71 \\
        ~ w/o Q-Connector & 37.43 & 78.66 & 21.21 & 26.39 \\
        ~ w/o D-Connector & 41.28 & 81.59 & 22.35 & 27.54 \\
    \bottomrule
    \end{tabular}
    \label{tab:ablation}
\end{table}

\begin{figure}[!t]
\vspace{0.1cm}
    \begin{subfigure}{.495\linewidth}
    \centering
    \includegraphics[width=\linewidth]{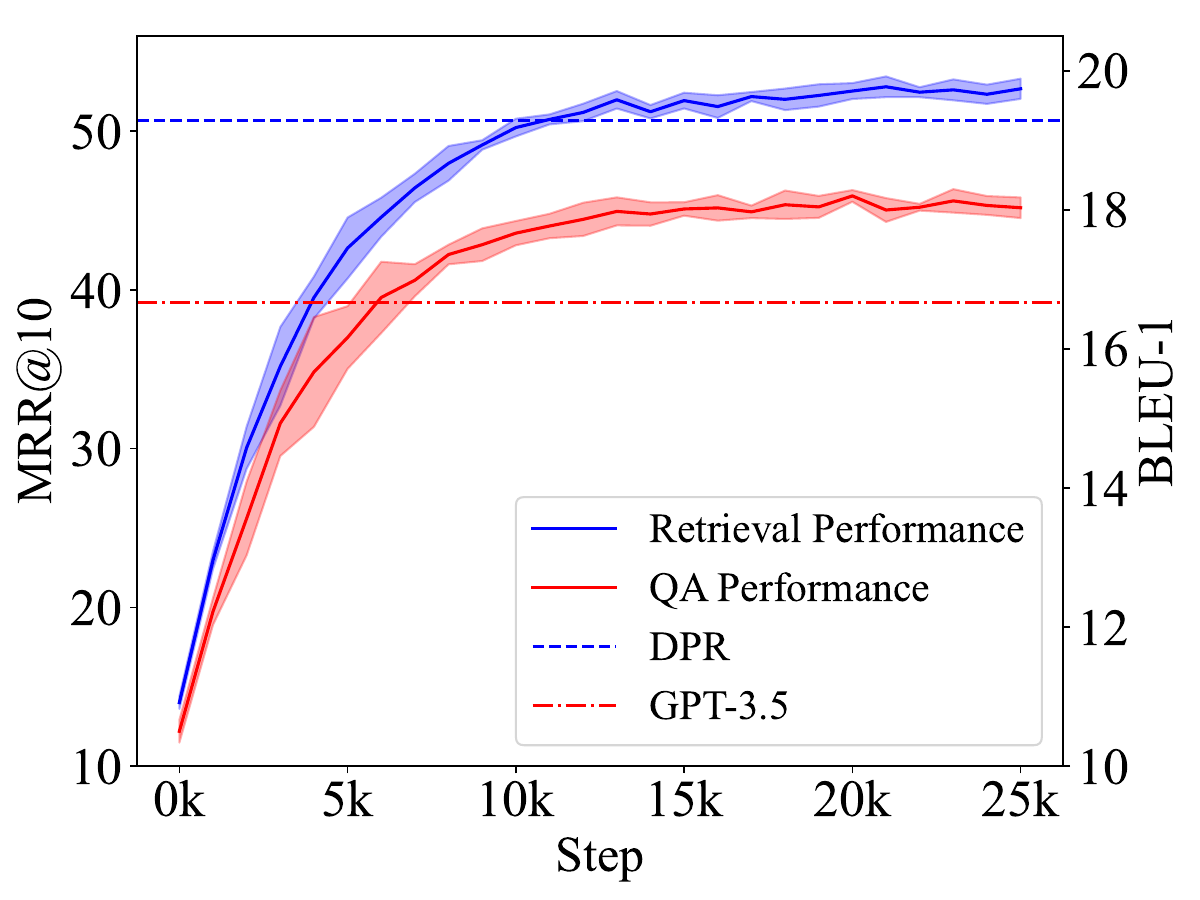}
    \caption{MS MARCO}
    \end{subfigure}
    \begin{subfigure}{.495\linewidth}
    \centering
    \includegraphics[width=\linewidth]{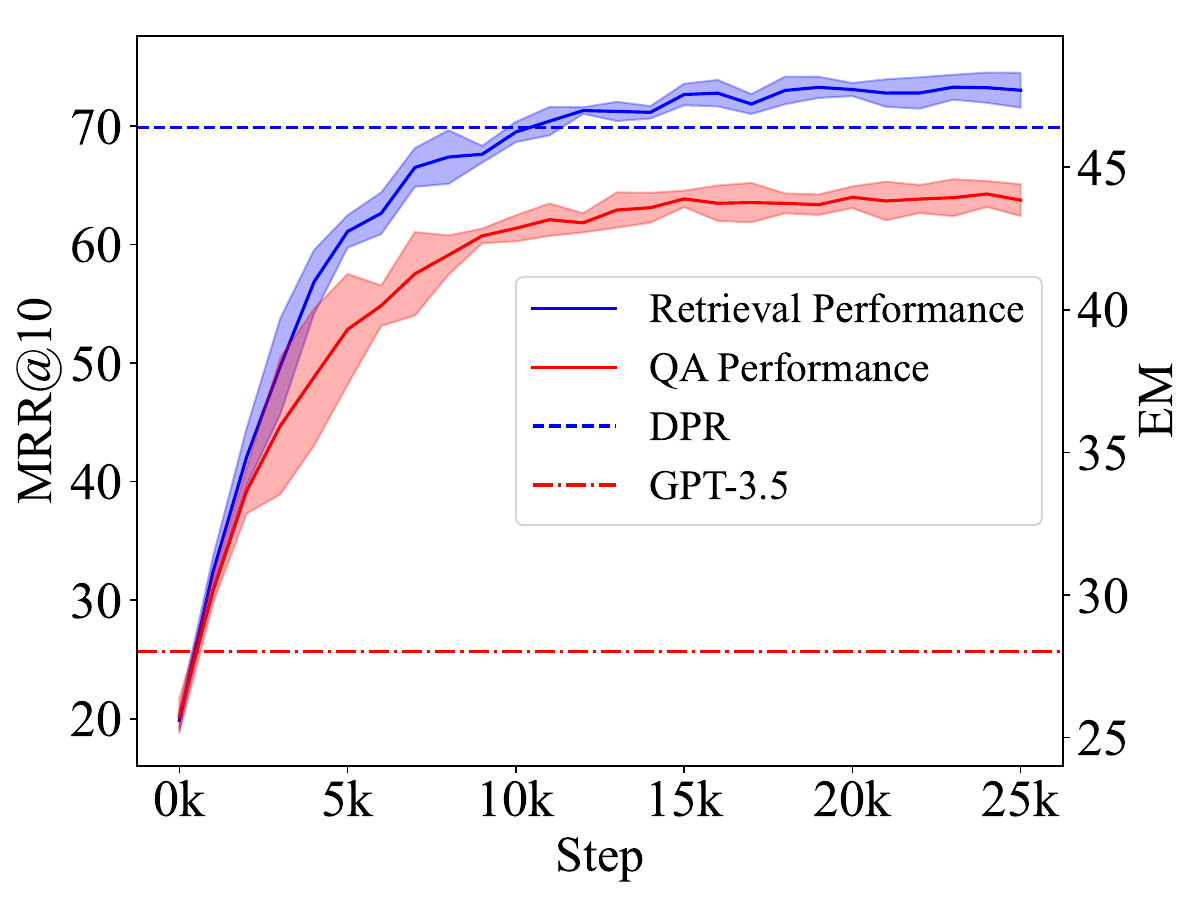}
    \caption{Natural Questions}
    \end{subfigure}
    \caption{Learning curves of retrieval and QA performance.} 
    \label{fig:learning_curve}
\end{figure}

\subsection{Study of Learning Curves}

To demonstrate the effectiveness of our proposed approach during the training process, we plot learning curves to showcase the retrieval and QA performance on the MS MARCO and NQ datasets. We utilize a combination of synthetic and labeled data to train the UniGen model. Figure \ref{fig:learning_curve} illustrates these curves, with the average values and standard deviations plotted for each metric, obtained from five separate training runs on each dataset.

The retrieval performance, measured by MRR@10, is represented by the blue curve, while the red curve represents the QA performance, measured by BLEU-1 for MS MARCO and EM for NQ. Notably, both tasks exhibit stable optimization throughout the learning process, thereby confirming the effectiveness of our proposed unified framework for simultaneous learning of retrieval and QA tasks.

\section{Conclusion}
In this paper, we present UniGen, a unified generative framework for retrieval and question answering. Our approach optimizes both tasks simultaneously and employs connectors generated by large language models to establish semantic connections in the input-output and docid-answer spaces. Additionally, our iterative enhancement approach proves to be effective in enhancing retrieval and QA performance. Through extensive experiments conducted on public datasets, we demonstrate the effectiveness of UniGen in both retrieval and QA tasks. This work opens up new possibilities for jointly learning retrieval and other generation tasks.

\section{Acknowledgement}
Zhicheng Dou is the corresponding author. This work was supported by the National Natural Science Foundation of China No. 62272467, Beijing Outstanding Young Scientist Program No. BJJWZYJH012019100020098, the fund for building world-class universities (disciplines) of Renmin University of China, and Public Computing Cloud, Renmin University of China. The work was partially done at Beijing Key Laboratory for Big Data Management and Analysis Methods.

\bibliography{reference}

\end{document}